\documentstyle[aps,prl,twocolumn,graphicx]{revtex}
\begin{document}
\draft % this command makes pacs numbers print

%-----------------------------------------------------------------------------

\title{Matrix controlled channel diffusion of sodium in amorphous silica}
\author{Emmanuel Sunyer, Philippe Jund$^\ast$  and R\'emi Jullien}

\address{
\vspace*{0.25cm}
Laboratoire des Verres - Universit\'e Montpellier 2\\
Place E. Bataillon Case 069, 34095 Montpellier France\\
\vspace*{0.25cm}
$^\ast$Laboratoire de Physicochimie de la Mati\`ere Condens\'ee - Universit\'e Montpellier 2\\
Place E. Bataillon Case 003, 34095 Montpellier France
}

\maketitle

\begin{abstract}
To find the origin of the diffusion channels observed in sodium-silicate
glasses, we have performed classical molecular dynamics simulations of
Na$_2$O--4SiO$_2$ during which the mass of the Si and O atoms has been
multiplied by a tuning coefficient. We observe that the channels disappear and
that the diffusive motion of the sodium atoms vanishes if this coefficient is
larger than a threshold value. Above this threshold the vibrational states of
the matrix are not compatible with those of the sodium ions. We interpret hence
the decrease of the diffusion by the absence of resonance conditions.
\end{abstract}


The mechanism of ionic transport in amorphous materials~\cite{intro,funke} at
the atomic scale is  not completely elucidated so far and therefore it is   
the subject of numerous  experimental~\cite{greaves,zotov_1} and
numerical~\cite{zotov_2,vessal,huang,oviedo,heuer} studies. Among the most
studied systems are the sodium-silicate glasses since they contain the
essential  ingredients, namely the amorphous matrix (silica) and the mobile
ions (sodium), as a first step in the simulation of more complex glasses of
higher  practical interest.

In previous studies~\cite{jund,sunyer_1,sunyer_2}, we have shown by means of
classical molecular dynamics simulations of Na$_2$O--4SiO$_2$ (NS4), that the
sodium atoms  diffuse through a well connected network of pockets (which
represents only a limited fraction of the entire available space) that we have
called ``channels'' to be coherent with the literature. The existence of the
channels, which are not due to micro-segregation effects~\cite{vessal,huang},
has been confirmed by the existence of a pre-peak in the partial Na-Na
structure factor at a wave-vector $q = 0.95$~\AA$^{-1}$\cite{sunyer_3}. This
pre-peak has also  been observed experimentally~\cite{meyer} and
numerically~\cite{horbach_2} in another study. We have also shown that the
location of the channels is strongly correlated  to the positions of the
non-bridging oxygens~\cite{sunyer_3} and Horbach \textit{et al.} have shown
that the sodium dynamics should be related to that of the underlying  silica
network~\cite{horbach_2}. This suggests that the origin of the channels could
be related to the dynamical properties of the matrix. To check this idea we
present in this letter classical molecular dynamics simulations on a series of
``toy'' systems in which the atomic masses of both the oxygen and silicon atoms
have been systematically changed after artificially multiplying their
experimental values by a common  factor $\mu$ varying from 0.5 to infinity, the
usual NS4 system being recovered for $\mu = 1$.

By studying both the mean square displacement of the sodium atoms and the
characteristics of the channels, we find a change in the sodium diffusion
properties when the parameter $\mu$ is increased above a value of about 30.
Above this threshold, the sodium diffusion decreases and the channels can no
more be clearly defined. Guided by the concomitant change in the short time
characteristics of the velocity autocorrelation function, we have calculated
the vibrational density of states (VDOS) for both the sodium atoms and the
atoms of the matrix for various values of $\mu$. We observe that the threshold
corresponds to the value of $\mu$ above which the sharp VDOS of the sodium
atoms starts to separate from the  band of vibrational states of the matrix. 
Therefore we propose that the diffusive motion of the sodium atoms,
\textit{i.e.} their ability to escape from their local cage, is facilitated
when their vibrational frequencies are compatible with the ones of the matrix.
We argue that this mechanism is responsible for the channel diffusion of sodium in
the silica matrix.

In this letter we present classical molecular dynamics calculations of a
system  of 648 particles (86 sodium, 173 silicon and 389 oxygen atoms) 
confined in a cubic box of edge length 20.88~\AA\ with periodic boundary
conditions. The density is thus the experimental density of glassy NS4,
\textit{i.e.} 2.38~g.cm$^{-3}$~\cite{bansal}. The interactions between the
particles are given by a  modified version of the so-called ``BKS''
potential~\cite{vanbeest,kramer} which is able to reproduce the structure as
well as the dynamics of several sodium-silicate systems~
\cite{jund,sunyer_1,sunyer_2,sunyer_3,horbach_1} 
(for more details see \cite{sunyer_1}). In the present study we have generated
three independent samples (in the following, all the results are averaged over
these samples in order to improve the statistics) at a temperature of $\sim
1900$~K for which the channels have been evidenced in our previous studies. For
each sample we  have performed ten different simulations in which the mass of
the atoms of the matrix has been artificially multiplied by a factor $\mu$ of 0.5,
1, 2, 5, 10, 30, 10$^2$, 10$^3$, 10$^4$, and 10$^6$. We have also carried out
the limiting case $\mu=\infty$ by performing simulations in which only the
sodium atoms move, the atoms of the matrix being kept fixed (frozen matrix
approximation). Each of these 33 samples has first been relaxed at $\sim
1900$~K  for 10$^6$ steps (1 step = 1.4~fs) and the following 10$^6$ steps were
used to  produce 1000 configurations recorded every 1000 time steps. These
configurations have then been used to analyze the trajectories of the sodium
atoms during these 1.4~ns.

In Fig.~1 we show $R^2(t)$, the mean square displacement (MSD) of the sodium
atoms, for various values of $\mu$. While these curves have the characteristic 
long-time shape of strong diffusion for small values of $\mu$, they flatten out
as $\mu$ increases. In the limiting case $\mu=\infty$ (frozen matrix) the curve
becomes even completely flat, at least within the period of time covered by
our  simulations. In order to give a quantitative idea of how the ionic
diffusion properties decrease with increasing $\mu$, we have determined a
characteristic time $\tau_{\rm MSD}$ necessary for $R^2(t)$ to be equal to
4~\AA$^2$. This corresponds to an average travel distance of 2~\AA\ which is
the {\em minimum} distance between two Na neighbors  (as determined from the
Na--Na radial pair distribution function \cite{sunyer_2}). In the inset of
Fig.~1 we give the variation of $\tau_{\rm MSD}$ with $\mu$. While $\tau_{\rm
MSD}$ is almost independent of $\mu$ for small $\mu$, it starts to increase 
for values of $\mu$ larger than a value of about 30.

Since we have previously shown that the diffusive motion of the sodium atoms
takes place within channels~\cite{jund,sunyer_1,sunyer_2,sunyer_3}, it is worth
studying how these channels are modified when the mass of the atoms of the matrix 
is changed. Therefore, for all our values of $\mu$, we have determined the
channels, in the same manner as previously~\cite{jund}: using a 3-dimensional
mesh we  determine the number of {\em different} sodium atoms that have visited
each cube of the mesh during the total simulation time. Then we define $\xi$
which is the minimal occupation number such that the cubes visited more than
$\xi$ times represent the upper 10~\% of all the visited cubes (for more
details see \cite{jund}). In previous studies we have shown that at $T \sim
1900$~K a cube needed to be visited by at least $\xi = 8$ different sodium  atoms
during a 1.4~ns simulation, in order to be part of the channel structure.

Here we study how $\xi$ changes with $\mu$ and the results (averaged over three
samples)  are reported in Fig.~2. While for $\mu < 30$, $\xi$ remains almost
constant between 7 and 8 within the error bars, it starts to decrease
dramatically for $\mu > 30$. Obviously, as explained in
refs.~\cite{jund,sunyer_1,sunyer_2,sunyer_3}, one can no more speak of
``channels'' if  $\xi$ becomes as small as unity. Therefore one can interpret
the results in Fig.~2 by assuming that the channels disappear progressively as
soon as $\mu > 30$. Then one can argue that the increase of the characteristic
time for the diffusion observed in Fig.~1 is intimately correlated with the
disappearance of the channels shown in Fig.~2.

In order to know if the observed change in the long time diffusive behavior  of
the sodium atoms is accompanied by a change in their short time dynamical
properties, we have calculated the Na velocity autocorrelation function
$\vartheta(t)=\left<\vec{v}(t_0)\vec{v}(t_0+t)\right>/\left<v^2(t_0)\right>$
and studied  its short time behavior. As seen in Fig.~3 the short time part of
$\vartheta(t)$ is characterized by a first minimum at $\tau_\vartheta$, typical
of the time for a sodium atom to bounce back and forth against the internal
boundaries of the cage in which it vibrates (well before being eventually able
to jump towards another cage). In the inset of Fig.~3 the intensity of the
minimum $\vartheta_{\rm m} = \vartheta(\tau_\vartheta)$ has been plotted as a
function of $\mu$. $\vartheta_{\rm m} \simeq 0.28$ for $\mu < 10$ while
$\vartheta_{\rm m}$ drops down to $-0.5$ for $\mu > 30$. This increasing depth
can be interpreted as an increasing stiffness of the internal cage boundaries
when $\mu$ is increased. In a classical mechanics picture, due to this
increased stiffness, the probability for a sodium atom to escape from its cage
becomes weaker and this is consistent with the vanishing diffusion observed in
Fig.~1.

To go further in our microscopical analysis, and to understand the changes
occurring around $\mu \sim 30$, we have calculated the VDOS of the sodium atoms
and the atoms of the matrix, by Fourier transforming the corresponding velocity
autocorrelation functions. It is known that such a method reproduces only
approximately the VDOS, since at finite temperature (here $T \sim 1900$~K),
the harmonic hypothesis does not fully hold but this approximation will not be
crucial for our arguments. In Fig.~4~(a) the VDOS of the oxygen atoms for
different values of $\mu$ are represented (a similar picture could be drawn for
the Si atoms). The VDOS for $\mu = 1$ is of course close to the one of
amorphous silica obtained experimentally except that it is less  structured,
which is a known drawback of the  BKS potential \cite{guillot}. Such a VDOS is
typical of a broad band of vibrational states, extending from $\nu = 0$ up to
$\nu_1 \simeq$ 35~THz, as a result of the  coupling between neighboring oxygen
and silicon atoms forming a strong random covalent network.  When increasing
the mass of the atoms of the matrix, one observes a shrinkage of the VDOS towards
low frequencies. The top of the oxygen band, $\nu_{\rm max}^{\rm O}$, has been
plotted versus $\mu$ in Fig~4~(c): as expected from standard solid state
physics   $\nu_{\rm max}^{\rm O}$ varies like $\nu_1 \mu^{-1/2}$. The situation
is quite different for the  Na VDOS represented in Fig.~4~(b) for three typical
values of $\mu$: it looks like a well defined peak centered at a frequency
$\nu^{\rm Na}_1$ close to 5~THz for small values of $\mu$. Such a peak can be
interpreted as being mostly due to the vibrations of the sodium  atoms in their
cage. Using the approximate formula for such a motion, $R^{-1}\sqrt{kT/m_{\rm
Na}}$, where $R = 1$~\AA\ is a typical size of the cage deduced from the MSD
and $m_{\rm Na}$ is the mass of a sodium atom, one finds a frequency of about 8~THz 
in reasonable agreement with $\nu^{\rm Na}_1$.
The broadening of the peak is then certainly due to the polydispersity of the
$R$ values as well as to the coupling of the sodium atoms with other species.
With increasing $\mu$ a second peak at $\nu^{\rm Na}_2$ grows out of the high
frequency part and increases while the peak at $\nu^{\rm Na}_1$ decreases and
shifts towards low frequencies. The variation of $\nu^{\rm Na}_1$ and $\nu^{\rm
Na}_2$  with $\mu$ is represented in Fig.~4~(c) where the main peak of the Na
VDOS is represented by the filled symbols. One sees that up to $\mu = 30$ the
peak at $\nu^{\rm Na}_1$ is the main peak while for higher values of $\mu$ the
peak at $\nu^{\rm Na}_2$ becomes the main peak. Once the transition is
fulfilled, the secondary Na peak follows the variation of $\nu_{\rm max}^{\rm
O}$ with $\mu$  while the principal peak remains at a constant frequency. Such
a behavior is typical of hybridization effects as commonly seen in electronic
and vibrational systems. It is due to the coupling between the sodium atoms and
the atoms of the matrix. The characteristic value $\mu_0$ for which the
principal peak in the Na VDOS starts to escape from the broad oxygen band can
be estimated by equating the top of the band $\nu_1 \mu_{0}^{-1/2}$ and the typical 
sodium frequency, 5~THz. This gives $\mu_0 \simeq 50$ (indicated by the arrow in Fig.~4~(c)), 
which is remarkably close to the value of 30 above which we have observed the change in 
the diffusive properties of the Na atoms.

Therefore the above analysis of the vibrational density of states provides a very
simple explanation of the predominance of channel diffusion for $\mu$ smaller
than $\mu_0$. In that case the sodium frequency peak lies within the limits of
the vibrational band of the matrix as shown in Fig.~4~(c). Therefore there
always exists a vibrational mode of the matrix with the same frequency as the
one of the sodium atoms. Using a simple picture in direct space, the
vibrational amplitudes of the sodium atoms can become very large due to
``resonance conditions'', giving them the possibility to escape from their cage
and to jump to a neighboring cage, as it is expected in a local picture of the
diffusive motion. Moreover, at the location where a sodium atom is connected to
an oxygen atom via a covalent-like bond, it is clear that the local
hybridization is important and this explains why the location of the channels,
\textit{i.e.} the preferential pathways for the sodium diffusive motion, is
strongly correlated to the position of the non-bridging oxygen atoms. On the
contrary, for $\mu > \mu_0$, the vibrational peak of the sodium atoms is
located outside the VDOS of the matrix (Fig.~4~(c)) and the resonance
conditions are more difficult to be fulfilled. As a consequence, the sodium
atoms stay much longer confined in their cages. This picture is consistent with
our interpretation of the increasing depth of the first minimum of the Na
velocity autocorrelation function, as being due to an increasing stiffness of
the internal cage boundaries when $\mu$ is increased above $\mu_0$.

In conclusion, we have demonstrated the essential role played by the vibrations
of the atoms of the matrix in the existence of the channel diffusion of sodium
inside an amorphous silica matrix. In a sense this is close to  the concept of
``matrix-mediated-coupling'' recently used to interpret the  mixed cation
effect in glasses \cite{ingram} except that we show here that the
\textit{direct} hybridization between the ionic modes and the modes of the matrix is
necessary to insure a fast diffusion process of the ions. This result is also
coherent with previous studies showing indirectly that the sodium dynamics
should be intimately linked to the one of the matrix~\cite{horbach_2}. Of
course it would be extremely interesting to extend the  present study to other
kinds of ions, such as Li, H, etc., to test the generality  of our
interpretation (according to preliminary results it appears indeed that a
similar behavior is observed for Li in SiO$_2$ \cite{heuer2}). This would 
constitute a real improvement in the understanding of the mechanisms of ionic 
transport in random media.

\bigskip

We thank Pr. K. Funke and Pr. A. Heuer for interesting discussions. Part of the
numerical calculations were done at the ``Centre Informatique National de
l'Enseignement Sup\'erieur'' (CINES) in Montpellier.

\begin{figure}
\vspace*{1cm}
\centerline{\includegraphics[width=10cm]{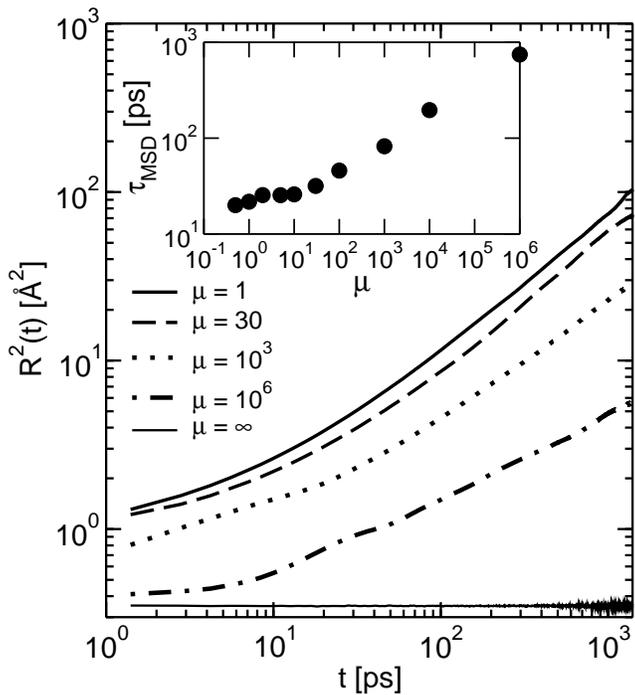}}
\caption{
Plot of the mean square displacement $R^2(t)$ of the sodium atoms for
different values of $\mu$. Inset: Plot of the characteristic diffusion time
$\tau_{\rm MSD}$ (see text for definition) as a function of $\mu$.
}
\label{fig1}
\end{figure}      
%

% figure 2 
\begin{figure}
\vspace*{1cm}
\centerline{\includegraphics[width=10cm]{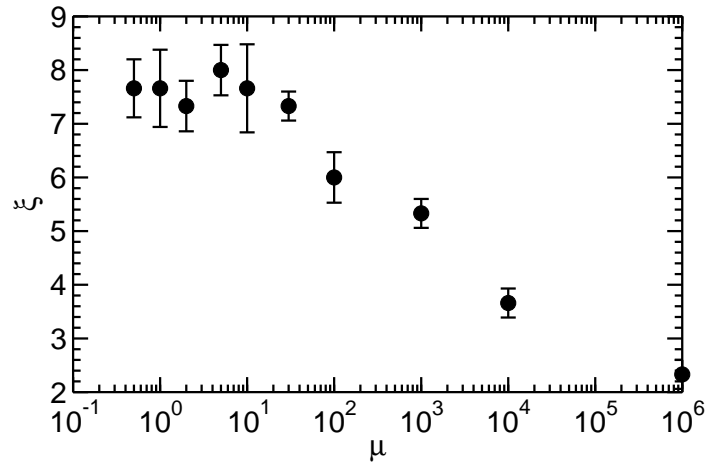}}
\caption{
Plot of $\xi$, the quantity used to define the channels (see text for further
details) as a function of $\mu$.
}
\end{figure}
%

% figure 3
\begin{figure}
\vspace*{1cm}
\centerline{\includegraphics[width=9cm]{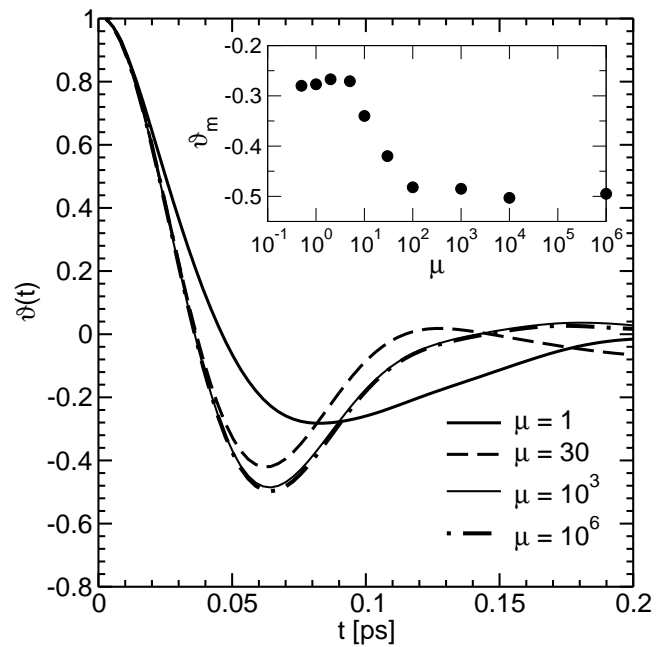}}
\caption{
Plot of the velocity autocorrelation function $\vartheta(t)$
 of the sodium atoms for different values of $\mu$. Inset: Plot of
the intensity of the first minimum $\vartheta_m$ as a function of $\mu$.
}
\end{figure}

\newpage
\onecolumn
% figure 4
\begin{figure}
\vspace*{1cm}
\hspace*{1cm}{\includegraphics[width=16cm]{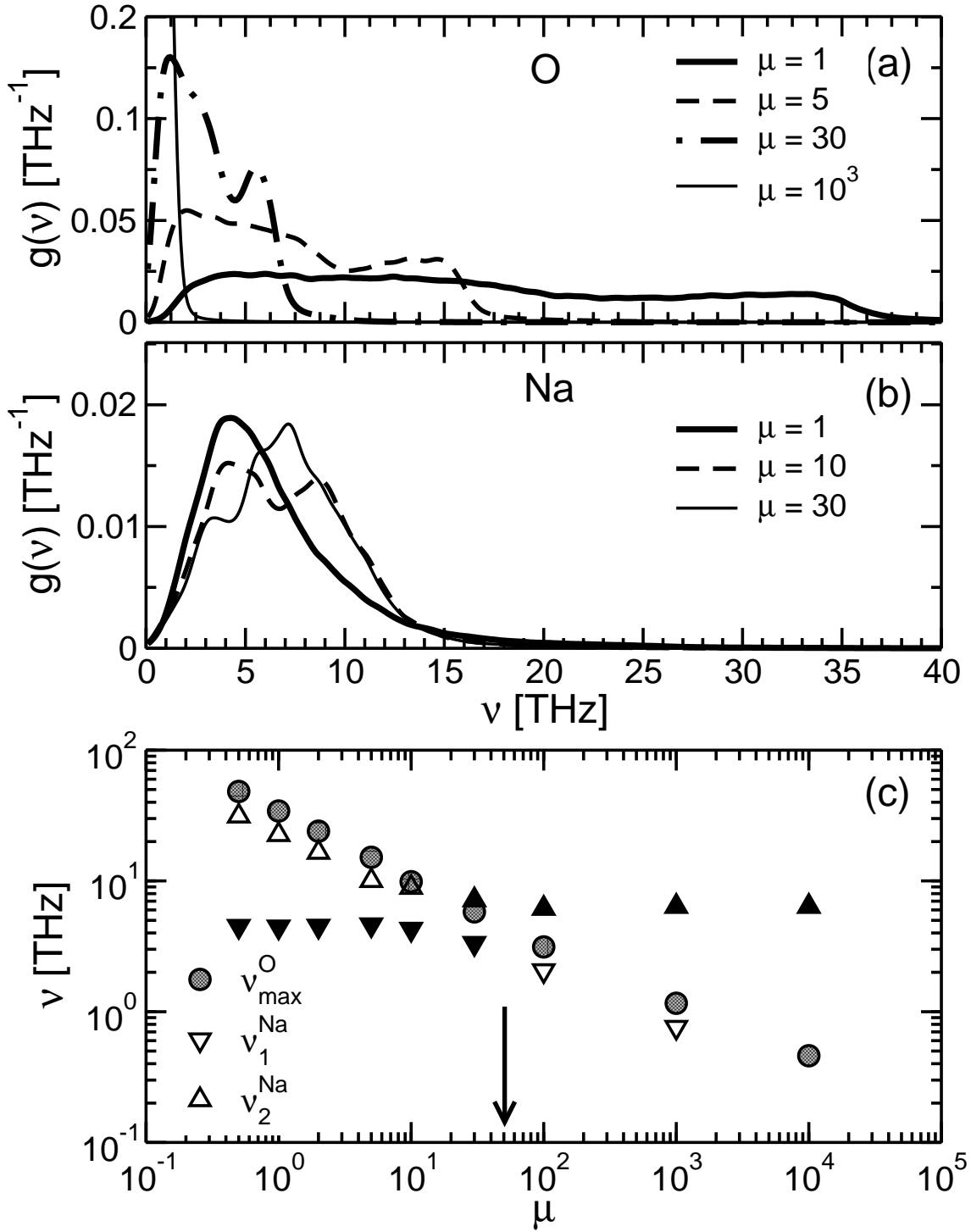}}
\caption{(a): Plot of the VDOS of the oxygen atoms for different values of $\mu$. (b): Plot
of the VDOS of the sodium atoms respectively for $\mu =1 $, 10, and 30. (c): Plot
of the highest frequency of the oxygen VDOS ({\Large$\bullet$}), the
frequency of the low energy peak ($\bigtriangledown$) and the high energy
peak ($\bigtriangleup$) of the sodium VDOS as a function of $\mu$. 
When these symbols are filled they indicate the main peak of the Na VDOS.
}
\end{figure}

\end{document}